\newcommand{\One}{{\bf 1}}

\newcommand{\dg}{\delta}
\newcommand{\sg}{\sigma}

\newcommand{\eg}{\epsilon}

\newcommand{\Cg}{\Gamma}
\newcommand{\Dg}{\Delta}
\newcommand{\vareg}{\varepsilon}

\newcommand{\di}{\partial}
\newcommand{\be}{\begin{equation}}
\newcommand{\ee}{\end{equation}}
\newcommand{\bearr}{\begin{eqnarray}}
\newcommand{\eearr}{\end{eqnarray}}

\newcommand{\Real}{{\bf R}}

\newcommand{\bfa}{{\bf a}}
\newcommand{\bfb}{{\bf b}}
\newcommand{\bfI}{{\bf I}}
\newcommand{\bfg}{{\bf g}}
\newcommand{\bfh}{{\bf h}}

\newcommand{\bfm}{{\bf m}}
\newcommand{\bfn}{{\bf n}}

\newcommand{\bfr}{{\bf r}}
\newcommand{\bfx}{{\bf x}}
\newcommand{\bfy}{{\bf y}}
\newcommand{\QED}{\rule{1mm}{3mm}}
\newcommand{\calD}{{\cal D}}

\newcommand{\calV}{{\cal V}}
\newcommand{\gothA}{{\bf A}}

\documentstyle[graphics,12pt]{article}

\begin{document}

\title{\Large \bf Spacetime as a Feynman 
diagram: 
the connection formulation\vskip.4cm}

\author{\normalsize \bf Michael P. Reisenberger\\
\normalsize \em 
Theoretical Physics Group, Blackett Laboratory,
\\ 
\normalsize\em Imperial College, London, SW7 2BZ, UK. \\ 
\normalsize\em Center for Gravitational Physics and Geometry, The Pennsylvania
\\ 
\normalsize\em   State University,  
104 Davey Lab, University Park, PA16802-6300, USA
\\
\normalsize\textsf{m.reisenberger@ic.ac.uk}\\[3mm]
\normalsize\rm \bf Carlo Rovelli\\
\normalsize\em Physics Department, University of Pittsburgh, \\ 
\normalsize\em Pittsburgh, PA 15260, USA.\\
\normalsize\em Centre de Physique Th\'eorique, 
CNRS Luminy, 13288 Marseille, France \rm
\\
\normalsize\textsf{rovelli@cpt.univ-mrs.fr }}
\maketitle

\begin{abstract}
\noindent Spin foam models are the path integral counterparts to loop
quantized canonical theories.  In the last few years several spin foam
models of gravity have been proposed, most of which live on finite
simplicial lattice spacetime.  The lattice truncates the presumably
infinite set of gravitational degrees of freedom down to a finite set. 
Models that can accomodate an infinite set of degrees of freedom and
that are independent of any background simplicial structure, or indeed
any {\em a priori} spacetime topology, can be obtained from the
lattice models by summing them over all lattice spacetimes.  Here we
show that this sum can be realized as the sum over Feynman diagrams
of a quantum field theory living on a suitable group manifold, with
each Feynman diagram defining a particular lattice spacetime.  We
give an explicit formula for the action of the field theory
corresponding to any given spin foam model in a wide class which
includes several gravity models.  Such a field theory was recently
found for a particular gravity model \cite{DPFKR}.  Our work
generalizes this result as well as Boulatov's and Ooguri's models of
three and four dimensional topological field theories, and ultimately
the old matrix models of two dimensional systems with dynamical
topology.  A first version of our result has appeared in a companion
paper \cite{Due}: here we present a new and more detailed derivation
based on the connection formulation of the spin foam models.
\end{abstract}

\section{Introduction}

The spin foam formalism \cite{Iwasaki,Rei94,RR,Rei97b,Baez98,Rov98,DP} 
has emerged in the last few years as an elegant 
synthesis of several approaches to quantum gravity and diffeomorphism 
invariant theories more generally. It can be viewed as a ``path integral" 
formulation corresponding to the canonical loop quantization framework 
\cite{Gambiniloop,RSloop,ALMMT,Zapata,spinnet} and also as an extension 
of the simplicial framework for topological field theories 
\cite{Ponzano,Turaev,Ooguri,CY}, which allows more 
general, non-topological, field theories to be represented.

Spin foams are coloured two dimensional complexes consisting of two 
dimensional faces (of arbitrary topology), joined on edges of valence 
$\geq 3$. Faces are coloured with non-trivial irreducible representations of 
a ``gauge group" $G$\footnote{
We will see that $G$ corresponds to the local gauge group or fibre bundle 
structure group commonly refered to as the ``gauge group" in discussions of 
Yang-Mills theory, as opposed to the group of all gauge transformations of 
the fields in spacetime.}
while edges carry ``intertwiners" - $G$ invariant tensors in the product 
representation formed by the representations on the incident faces. In a 
spin foam model the spin foams are regarded as histories of the physical 
system and assigned quantum amplitudes. In models of gravity a spin foam 
defines a discrete spacetime geometry \cite{Rei94,Ponzano,Rovelli93}; Spin 
foam models of gravity incorporate the discreteness of the geometry of space 
first uncovered \cite{discarea,discvol,DParea,ALarea,Thiemann_length} in loop 
quantized 
canonical theory into a spacetime sum over 
histories formalism naturally suited to a 4-diffeo invariant theory. This is 
very appealing because the discreteness, which is not ``put in by hand" but 
arises naturally in this framework,\footnote{
In using the spin foam formalism, or the canonical loop quantization formalism
one is making an ansatz, so it is possible that the discreteness is not real 
because the whole {\em framework} is not the one used by Nature. In lattice 
spin foam models there is another possibility, namely that an infinite 
renormalization of the bare theory that is necessary to go to the continuum 
limit wipes out the discreteness of the geometry in this limit. Finally it 
should be noted that in a spin foam model of Lorentzian 2+1 general relativity 
developed by Freidel \cite{Freidel} geometrical observables do not have an 
entirely discrete spectrum but there is a non-zero minimal length. This is a 
result of the non-compactness of the three dimensional Lorentz group $SO(2,1)$
and might be a general feature of Lorentzian models.} 
promises to remove the ultraviolet divergences found in perturbative theories 
of quantum gravity and matter fields.

A number of spin foam models of Euclidean\footnote{
These models are not Euclidean field theories in the ordinary sense because
essentially the exponential of {\em i} times the action is used to weight 
histories. We call these models of Euclidean quantum gravity because they are 
proposals for quantizations of classical general relativity with metric 
signature $++++$.}
\cite{RR,Rei97b,Rei97a,BC,Iwasaki99} and Lorentzian 
\cite{MS,Gupta,BCLor} quantum gravity in four spacetime dimensions have been 
proposed. Three dimensional general relativity \cite{Ponzano,Iwasaki,Freidel}
as well as four dimensional BF theory \cite{Ooguri,CY,Rei94}
can also be given a spin foam formulation. Even hypercubic Yang-Mills 
theory can be expressed in this framework. In fact any lattice model in which 
a connection with a compact gauge group forms the boundary data can be 
translated into a spin foam model \cite{Rei97b}.

Of the spin foam models of gravity referred to above all but \cite{RR} 
and \cite{Gupta} are simplicial lattice models. A shortcoming of such models 
is that their predictions depend on the simplicial complex chosen to represent
spacetime. Topological field theories, such as three dimensional general 
relativity or BF theory, can be formulated on a triangulated manifold in such 
a way as to be independent of the particular 
triangulation of the manifold chosen, essentially because such theories have 
only a finite number of degrees of freedom associated to global topological 
features of the manifold. Four dimensional gravity models on the other hand 
are expected to have infinitely many degrees of freedom in any given spacetime 
topology, so any finite triangulation (in which each simplex carries a finite 
set of degrees of freedom as in the proposed models) necessarily truncates and
thus misrepresents the gravitational field. 

Some sort of continuum limit must be taken. One approach is to sum over 
triangulations of the spacetime manifold. This seems very difficult to do in 
three or more dimensions since it is difficult to identify the topology of a 
given simplicial complex. A technically easier, but also more ambitous 
approach would be to sum over all simplicial manifolds\footnote{
Simplicial manifolds are simplicial complexes that are also manifolds. To be a 
manifold a complex must satisfy several local requirements the most 
non-trivial of which is that the boundary of the union of the simplicies 
incident on a point is a sphere.}
and thus sum over both triangulations {\em and} spacetime topology. It is an 
old dream of gravity theorists to make topology dynamical, thus removing yet 
another element of {\em a priori} background structure. It would be 
interesting to sum even models formulated on a continous spacetime manifold, 
such as those found in \cite{RR,Rov98} and \cite{Gupta}, over spacetime 
topologies.

In the present paper we will consider another, even wider, summation. 
Simplicial spin foams live on the 2-skeleton of the cellular complex dual to 
the simplicial complex. In other words, the spin foam faces consist of dual 
2-cells, and the spin foam edges of dual 1-cells. The dual 2-skeleton, the two 
dimensional complex formed by these cells (together with the dual 0-cells), 
thus {\em is} spacetime as far as spin foams are concerned.
We therefore propose to sum spin foam models of four spacetime dimensional 
gravity over all ``admissible" 2-complexes - 2-complexes that preserve some 
simple local features of dual 2-skeletons of triangulated 4-manifolds. 
Specifically we will require of admissible 2-complexes that their
0-, 1-, and 2- cells are topologically points, line segments, and disks
respectively. Furthermore we require that the combinatorial structure at each 
0-cell is that of a dual 2-skeleton to a triangulated 4-manifold. 
Thus each 0-cell has incident on it five endpoints of 1-cells and ten corners 
of 2-cells with each incident 2-cell corner bounded by two of the incident 
1-cells. This corresponds to the fact that a 4-simplex (dual to a 0-cell)
has in its boundary five 3-simplices (dual to 1-cells) and ten 2-simplices 
(dual to 2-cells).
\begin{figure}
\begin{center}
\resizebox{4in}{!}{\includegraphics{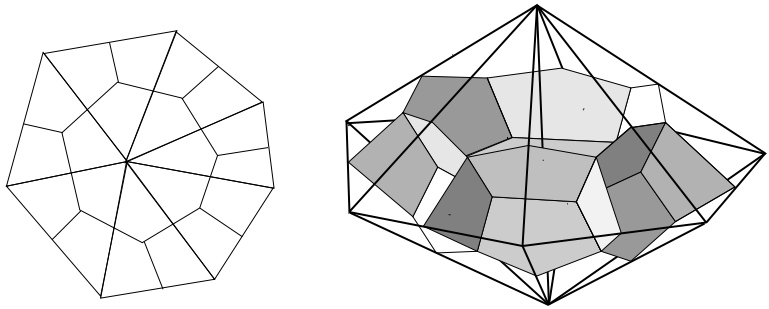}}
\vskip.4cm
\caption{This figure illustrates simplicial complexes and their dual skeletons
in two and three dimensions, as an aid to the reader contemplating the
the four dimensional simplicial complexes and their dual 2-skeletons that 
appear in our discussions. 
On the left a complex of seven triangles (2-simplices) along with 
its dual 1-skeleton is shown. On the right a complex of six 3-simplices 
(sharing a common 1-simplex) and its dual 2-skeleton is shown, with the dual 
2-cells shaded in. In both cases the dual skeleton is cut off at the 
boundary of the simplicial complex.}
\label{duals}
\end{center}
\end{figure}
(See Fig. \ref{duals}).
In a sufficiently small neighborhood of a 0-cell the 2-complex is 
indistinguishable from a dual 2-skeleton of a 4-d simplicial complex. 
Note however that we do not forbid 1-cells and 2-cells 
from being incident several times on the same 0-cell, so some fairly strange 
2-complexes are included in the sum.

Our main result is that this summation of a spin foam model over 2-complex 
spacetimes may be realized as the sum over Feynman diagrams in a 
perturbation expansion of a quantum field theory: The spin foam model defines 
a (somewhat unusual) quantum field theory on the Cartesian product of four 
copies of the gauge group $G$ the Feynman diagrams of which are precisely the
admissible 2-complexes, with the amplitude of each given, modulo symmetry
factors, by the spin foam sum on that 2-complex. (Comparing with a familiar 
QFT such as scalar $\lambda\phi^4$ theory in Minkowski space we see that the 
spacetime 2-complex plays the role of the graph of a Feynman diagram, while 
the spin foams living on the the 2-complex play the roles of the values of 
the momenta on the edges of a Feynman diagram, which are integrated over to 
obtain the amplitude of the diagram). We propose to adopt the 
perturbation series of the field theory, with its symmetry factors, as the 
definition of the spin foam model summed over 2-complex spacetimes.

To make things more concrete let's consider the amplitude for given boundary 
data on the boundary of a spacetime region. This amplitude is the sum of the 
amplitudes of the histories that match the boundary data.\footnote{
This amplitude can be thought of as the quantum probability amplitude of the 
boundary data in the Hartle-Hawking vacuum. Alternatively, if the boundary is 
divided into past and future parts then the amplitude can be interpreted
as the transition amplitude from past to future data.}
In a spin foam model on a fixed 2-complex $J$ representing spacetime the 
boundary is a graph $\Cg = \di J$ and the sum runs over all spin foams living 
on $J$ matching the boundary data - irreducible representations on the edges of
$\Gamma$ and intertwiners on the nodes. Now we wish to sum the amplitude 
obtained by summing over spin foams also over all $J$ bounded by the same, 
fixed, $\Gamma$,\footnote{
In the sum we are contemplating $\Gamma$ need not be the boundary of $J$ in 
quite the usual sense. We allow the possibility that parts of $\Gamma$ are 
glued together instead of to $J$. The situation is similar to that of 
encountered with shrink wrapped foods in which
part of the plastic covering is stuck to itself rather than the food item. Returning to a more
theoretical setting, this peculiar feature of the boundaries we admit is physically quite natural.
If we divide $\Gamma$ into a future half $\Gamma_+$ and a past half $\Gamma_-$ and interpret the 
amplitude of the boundary data as the transition amplitude from the past to the future boundary 
data, then there is no reason to exclude complexes representing ``bubble" time evolution
in which parts of $\Gamma_+$ and $\Gamma_-$ coincide, and no reason to exclude such complexes from
a sum over complexes interpolating between $\Gamma_+$ and $\Gamma_-$.}
 which is to be 
considered part of the boundary data. In the field theory picture 
this sum over 2-complexes is the Feynman diagram expansion of the expectation value
of an observable that encodes the boundary data. The field theory formulation
provides us with formal functional integral expressions for this expectation value,
and any other quantitiy of interest, which are often easier to manipulate than the 
original sums. Moreover, a regularization scheme that renders well defined the functional 
integral would also provide a definition of the values of the sums. 
We shall return to this point.

Our work generalizes the results of Ooguri \cite{Ooguri} and De Pietri {\em et. al.} 
\cite{DPFKR}
who obtained field theory formulations for two particular models (summed over spacetime
2-complexes), namely $SU(2)$ BF theory \cite{BF} and the Barrett-Crane
model \cite{BC} of Euclidean quantum gravity. Many of the underlying ideas go back to the
work of Boulatov \cite{Bou}, on three dimensional gravity, and to matrix models of two dimensional
theories \cite{matrix}.

The remainder of the paper is organized as follows. In \S \ref{field_theory} we
define precisely the class of spin foam models under consideration and then 
we present and verify our main result: an explicit formula for the action of 
the field theory that defines the sum over 2-complexes of any given spin foam 
model in this class. The Turaev-Ooguri-Crane-Yetter model of BF theory
is discussed as an example. Throughout this section spin foam models  
are formulated as lattice gauge theories on a special type of lattice.
In \S \ref{spin_foam} we show how these spin foam models, and the field theory
generating the sum over spacetime 2-complexes, are represented in terms
spin foams. The issue of divergences, regularization and renormalization is
also briefly touched on. In the last section we give an argument
suggesting that the field theory is actually finite for regulated spin foam 
models (which is not at all obvious from the perspective of a sum over
spacetime 2-complexes) and outline how the theory can be extended to 
accomodate more general spin foam models.

\section{Field theory formulation of spin foam models with dynamical topology}

\label{field_theory}

\subsection{Local spin foam models in a lattice connection formulation} 

In the present section we obtain, explicitly, a field theory defining the sum 
over 2-complex spacetimes for a wide class of ``local" spin foam models living 
on the dual 2-skeletons of triangulated 4-manifolds, or more generally on 
admissible 2-complexes. This class includes all Euclidean simplicial four 
dimensional gravity models except that of Iwasaki \cite{Iwasaki99}.

Instead of working directly with spin foam sums we will use the connection 
formulation of the spin foam models because this provides the easiest route to 
our result. In \S \ref{spin_foam} we will show how the result looks in the 
spin foam formulation and indicate how it may be obtianed 
within that framework. In the connection formulation a history consists of a 
connection specified by elements of $G$ defining parallel transport along 
``boundary edges" which run from the center of a 2-cell of $J$ to the center 
of one of its bounding 1-cells.\footnote{
Where the centers of these cells are placed does not really matter. In fact the parallel transporters
could simply be associated with pairs consisting of a 2-cell and a 1-cell in its boundary.}
\begin{figure}
\begin{center}
\resizebox{4in}{!}{\includegraphics{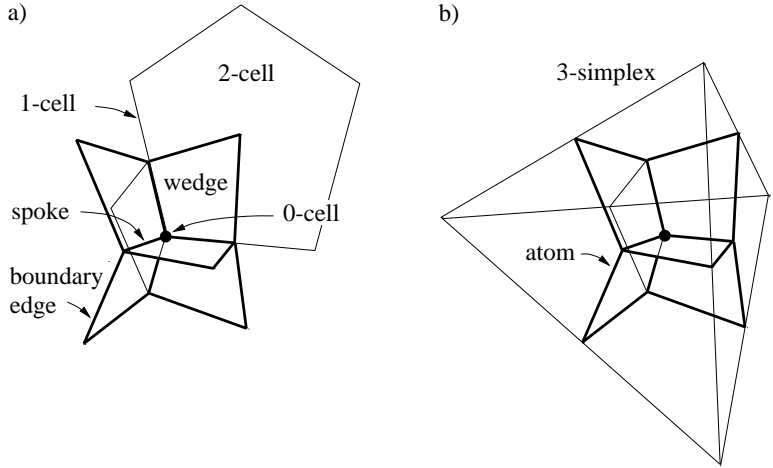}}
\vskip.4cm
\caption{Atoms and their relation to simplices are illustrated in three
dimensions. Panel a) shows an atom and one complete dual 2-cell of which the
atom has a wedge. Panel b) shows the 3-simplex inside which the atom would 
live were it part of a dual 2-complex of a three dimensional simplicial 
complex. The ``four dimensional'' atoms that we deal with in the text differs
from that shown in that it is identical to the portion inside a 
{\em 4-simplex} of the 2-skeleton dual to a {\em four} dimensional simplicial 
complex. This means that the atom has five one dimensional cells and ten
wedges, as opposed to the four one dimensional cells and six wedges shown 
here.}
\label{atomfig}
\end{center}
\end{figure}
As illustrated in Fig. \ref{atomfig} a) these edges cut $J$ into disjoint, 
topologically identical pieces which we shall call ``atoms", because they can 
be viewed as the fundamental building blocks from which any admissible 
2-complex can be built. Each atom contains one 0-cell, five one dimensional 
cells called ``spokes'' (portions of 1-cells of $J$), and ten two 
dimensional cells called ``wedges" (portions of 2-cells of $J$). It is 
bounded by a graph made of boundary edges that is homeomorphic 
to the 1-skeleton $\Gamma_5$ of a 4-simplex, having five 4-valent nodes 
connected in all possible ways by chains consisting of two boundary edges. 
The 4-valent 
nodes are the centers of incident 1-cells while the 2-valent nodes in the 
chains are the centers of incident 2-cells. (Henceforth when we 
speak of the nodes of the atomic boundaries we shall mean only the 4-valent 
nodes unless the 2-valent nodes are explicitly included). When $J$ is the 
dual 2-skeleton of a four dimensional simplicial complex the boundary edges are
precisely the edges where the boundaries of the 4-simplices cut the dual 
2-skeleton, and the atoms that they cut $J$ into are the portions of $J$ 
inside each 4-simplex. Atoms are in this sense the dual 2-skeletons of 
4-simplices. See Fig. \ref{atomfig} b).

The models we shall consider have compact $G$ and they are local. Each is 
defined by a ``vertex function" $V$, a gauge invariant function of a 
connection on $\Gamma_5$. This function evaluated on the connection on the 
boundary of an atom gives the quantum 
mechanical amplitude for that connection. The amplitude for the whole connection on all of
$J$ is then the product of the amplitudes for all of the atoms.

If we number the 4-valent nodes of $\Gamma_5$ from 1 to 5 and let the indices $i,j,k,...$ range over
these numbers then we may indicate the oriented edge of $\Gamma_5$ from node $i$ to node $j$
by $l_{ij}$ and the half edge from $i$ to the center of $l_{ij}$ (which corresponds to a boundary
edge in the boundary of an atom) by $e_i{}^j$. $V$ is then a function of the parallel transporters
$g_{ij} = h_i{}^j [h_j{}^i]^{-1}$ along the $l_{ij}$, where $h_i{}^j \in G$ is the parallel transporter
along $e_i{}^j$. In the model the amplitude for one atom, $x$, is thus 
$V(h_{x\,i}{}^j [h_{x\,j}{}^i]^{-1})$ where $h_{x\,i}{}^j$ is the parallel 
transporter along the boundary edge mapped to $e_i{}^j$, and
the amplitude for the connection on the whole spacetime is
\be	\label{weight}
w = \prod_{x \in\: \mbox{\scriptsize atoms of}\ J} V(h_{x\,i}{}^j [h_{x\,j}{}^i]^{-1}).
\ee
(Since each atom contains one 0-cell the product can also be viewed as a 
product over 0-cells of $J$).\footnote{
A footnote on vector and matrix notation:
In (\ref{weight}) $h_{x\,i}^j$ indicates the whole matrix of group
elements $[h_{x\,i}^j]_{1 \leq i < j \leq 5}$ in a sort of abstract index 
notation. It should be clear from the context when such expressions denote
a whole matrix or vector or just one element. On other occasions we will use
boldface letters to denote matrices or vectors, so the matrix of 
$h_{x\,i}{}^j$s would be written as $\bfh_x$ and the second row of this matrix
can be written as $\bfh_{x\,2}{}^.$. Again information that is clear from the 
context will be left out, with $\bfh$ in one context refering to all $h$s in 
the whole 2-complex while in another only to the four $h$s associated
with a given node on an atomic boundary. Finally we shall use the 
Einstein summation convention on gauge group tensor indices, so any repeated 
index of this type is summed over.}

If the model is a simplicial approximation to a spacetime manifold field theory, as all the gravity
models aim to be, then $V$ is an approximation to $\exp(i[\mbox{Effective action in 4-simplex}])$
in terms of the connection on the boundary of the atom in the 4-simplex. The atomic boundary is a 
graph in the boundary of the simplex so the connection on it is a discrete approximation
to the continuum connection on the boundary of the 4-simplex.

As is proper for a local theory regions of spacetime (composed of whole atoms) communicate
only via boundary data - the connection on the boundary. A model in this connection formalism
is nothing other than a lattice gauge theory defined on an unusual lattice in order to bring 
boundary data and locality to the fore. For this reason
this class of models was called ``local lattice gauge theory" in \cite{Rei97b}.


\subsection{The main result: the field theory that generates the sum over 
topologies}

We can now introduce the field theory that generates the sum over admissible 2-complexes of
the model defined by the vertex function $V$. It is a real scalar field theory on $G^4$, 
the Cartesian product of four copies of the gauge group manifold, determined by the action
\be		\label{action_def}
I[\psi] = I_0[\psi] - \lambda \calV[\psi],
\ee
where
\be		\label{I0_def}
I_0[\psi] = \frac{1}{2\cdot 4!} \int_{G^4} d^4h\ \psi^2(h_1,h_2,h_3,h_4)
\ee
and
\be		\label{calV_def}
\calV[\psi] = \frac{1}{5!}\int_{G^{20}} d^{20}h\ V(h_i{}^j [h_j{}^i]^{-1})\:\psi(h_1{}^i)
\psi(h_2{}^i)\psi(h_3{}^i)\psi(h_4{}^i)\psi(h_5{}^i).
\ee
$I_0$ is the kinetic term. It is quadratic, but it contains no derivatives. $\lambda \calV$
is a 5th order, non-local interaction term, with $\lambda$ the coupling constant in which we will
expand to get the perturbation series. The scalar field $\psi(h_1,h_2,h_3,h_4)$
is required to be symmetric in its four arguments.

A wavefunction $\theta$ of the connection on boundary $\Gamma$ is represented by the 
observable\footnote{
The observable, mentioned in the introduction, that represents a particular boundary connection
$\bfh_\Gamma$ is obtained by taking $\theta$ to be a delta distribution on the gauge equivalence
class of $\bfh_\Gamma$.}
\be		\label{Theta_def}
\Theta[\psi] = \frac{1}{sym(\Gamma)} \int d\bfh_\Gamma\ \theta(\bfh_\Gamma) \prod_{a\in 
\mbox{\scriptsize nodes of}\ \Gamma} \psi(h_a{}^b).
\ee
The arguments $h_a{}^b$ of $\psi$ are the parallel transporters along the boundary edges 
$e_a{}^b$ from the node $a$ toward the four neighboring nodes (indexed by $b$);
$\bfh_\Gamma$ represents the whole connection on $\Gamma$, i.e. the vector of all the
parallel transporters along boundary edges in $\Gamma$; and $sym(\Gamma)$ is the number
of symmetries of $\Gamma$, i.e. the number of mappings of the set of nodes of $\Gamma$
to itself that preserve the adjacency matrix. (Since the identity mapping is a symmetry
$sym(\Gamma) \geq 1$).

The main result of the present paper is the following:
\newline

\noindent {\bf Theorem}

The formal perturbation series of the expectation value of $\Theta^*$,
\be		\label{VEV}
\int \calD\psi\ e^{-I_0[\psi] + \lambda \calV[\psi]}\ \Theta^*[\psi],
\ee
in powers of $\lambda$ is
\be	\label{Feynman_sum}
\sum_{J \in \gothA_\Gamma} \frac{\lambda^{n(J)}}{sym(J)} \int d\bfh\ \theta^*(\bfh_\Gamma) w(\bfh),
\ee
the sum over the set $\gothA_\Gamma$ of admissible 2-complexes $J$ bounded by $\Gamma$ of the
overlap of the state $\theta$ with the Hartle-Hawking vacuum of the spin foam model on $J$.
$n(J)$ is the number 0-cells in $J$ and $sym(J)$ is the number of symmetries of $J$, i.e. of mappings
of $J$ to itself that preserve the combinatorial structure of $J$.\footnote{
The identity mapping is included in this set so $sym(J)$ is at least 1.}
$\calD\psi$ is normalized so that $\int\calD\psi\ e^{-I_0[\psi]} = 1$.
\newline

\noindent {\em Proof:} The proof is similar to the derivation the Feynman diagram expansion 
in local field theories in Minkowski space.
The order $\lambda^n$ term in (\ref{VEV}) is 
\be		\label{order_n}
\frac{\lambda^n}{n!} \int \calD\psi\ e^{-I_0[\psi]}\ \calV^n[\psi]\Theta^*[\psi].
\ee
The functional integral is Gaussian. It's value is zero if the number of factors of $\psi$ in
$\calV^n\Theta^*$ is odd. If the number is even the value of the integral is a sum of terms
associated with each possible partition of the factors of $\psi$ in $\calV^n\Theta^*$ into pairs.
Here we imagine that $\calV^n\Theta^*$ has been written as $n$ factors of $\calV$ followed by
$\Theta^*$ and that the $\calV$s and $\Theta^*$ have been written explicitly as the integrals
(\ref{calV_def}) and (\ref{Theta_def}) with the factors of $\psi$ in the integrals appearing in 
some definite order. The pairings being summed over are the distinct pairings of the elements
in the ordered sequence of $\psi$s appearing in this explicit expression for $\calV^n\Theta^*$.
The term corresponding to a given pairing is obtained by replacing each pair 
$\psi(\bfh)\psi(\bfh')$ by the corresponding propagator
\bearr
\langle \psi(\bfh)\psi(\bfh')\rangle_0 & \equiv & \int \calD\psi 
\ e^{-I_0[\psi]}\ \psi(\bfh)\psi(\bfh')  \\
& = & - \int \calD\psi\ \psi(\bfh) \sum_{\sg\in S_4} \frac{\dg}{\dg\psi(\sg[\bfh'])}
e^{-I_0[\psi]} 			\label{deriv} \\
& = & \sum_{\sg\in S_4} \int \calD\psi \frac{\dg\psi(\bfh)}{\dg\psi(\sg[\bfh'])}
e^{-I_0[\psi]} \\
& = & \sum_{\sg\in S_4} \dg(h_1,h'_{\sg(1)})\dg(h_2,h'_{\sg(2)})\dg(h_3,h'_{\sg(3)})
\dg(h_4,h'_{\sg(4)}).	\label{propagator}
\eearr
Here $\bfh$ and $\bfh'$ are the sequences of four group element arguments of $\psi$, and
$\sg[\bfh'] = [h'_{\sg(1)},...,h'_{\sg(4)}]$ is $\bfh'$ reordered according to the permutation
$\sg$. Note that the symmetrization had to be introduced in (\ref{deriv}) because the integral
runs over symmetrized $\psi$ only. Thus the integration by parts is justified only for the 
symmetrized derivative.\footnote{
Note also that the delta distributions in (\ref{propagator})
are normalized Haar measure deltas. The definition (\ref{I0_def}) of $I_0$ is a Haar 
measure integral and the functional derivatives in (\ref{deriv}) are Haar 
measure functional derivatives.}

$\calV^n \Theta^*$ is an integral over the connections on the boundaries of $n$ separate
atoms and on $\Gamma$. When each of the pairs of $\psi$s is replaced by a particular term
in the sum (\ref{propagator}) for the propagator the delta 
distributions reduce the integral to one over {\em matching} connections 
on the $n$ atomic boundaries and $\Gamma$, where in each term the nodes are matched up 
according to the pairing of the $\psi$s and the boundary edges attached to each node 
are matched according to the permutation $\sg \in S_4$ of the particular term of (\ref{propagator})
being employed. (Actually, to make the matching of boundary edges corresponding to a given 
$\sg$ unambigous we need to adopt a convention to specify the {\em unpermuted} order of
these edges on the atoms and on $\Gamma$. Since boundary edges $e_a{}^b$ are indexed by pairs
of nodes $a,b$ the ordering of the $\psi$s, which is equivalent to an ordering of the
nodes, provides a natural choice for the ordering of the boundary edges attached to a 
particular node $a$.) 

Glueing together the atoms and $\Gamma$ in this way produces an admissible 2-complex $J$
with boundary $\Gamma$. The corresponding contribution to (\ref{order_n}) is 
\be		\label{term}
\frac{\lambda^n}{n!} (\frac{1}{5!})^n \frac{1}{sym(\Gamma)} \int d\bfh\ \theta^*(\bfh_\Gamma) w(\bfh).
\ee
To evaluate (\ref{order_n}) we must add up the contributions (\ref{term}) corresponding to
each glueing, i.e. to each to each pairing of $\psi$s with permutations $\sg \in S_4$ associated
to each pair. Since (\ref{term}) depends only on the structure of $J$ we can first sum over
glueings that lead to the same $J$ by multiplying (\ref{term}) by the number of such glueings,
and then sum over $J$s.

In our expansion the atoms and the nodes in each atomic boundary as well as in $\Gamma$ are
numbered, in an order arising from the way in which the (\ref{order_n}) was written
explicitly as an integral of a product of $\psi$s. When the atoms and $\Gamma$ are glued together
to form a 2-complex $J$ this reference numbering defines a numbering of the atoms, atomic boundary 
nodes and $\Gamma$ nodes of $J$. Conversely, if any admissible, $n$ atom 2-complex bounded by 
$\Gamma$ is equiped with a ``good" numbering then it defines uniquely a glueing of the
atoms and $\Gamma$ with their reference numbering, and thus a pairing of the sequence 
of $\psi$s used in the expansion and a set of associated $S_4$ permutations. Here a good numbering 
is an arbitrary numbering of the atoms and of the nodes in each atom together with a numbering of the
nodes of $\Gamma$ such that $\Gamma$ has the same adjacency matrix as in the reference numbering.
From this we conclude, firstly, that our expansion generates {\em all} admissible, $n$ atom
2-complexes bounded by $\Gamma$, so the sum should run over all such complexes, and secondly,
that all glueings that lead to the same 2-complex arise from good numberings of the atoms and nodes
in that 2-complex. 

There are $n!(5!)^n sym(\Gamma)$ good numberings, but renumbering does
not {\em necessarily} produce a new glueing. A good renumbering of $J$ that produces the same 
gluing is a symmetry of $J$, because it defines a mapping of the set of cells of $J$ to itself 
that preserves all incidences, and conversely.  
There are thus $n! (5!)^n sym(\Gamma)/sym(J)$ glueings that produce the complex $J$ so
(\ref{order_n}) equals
\be		\label{J_sum}
\sum_{\{J \in \gothA_\Gamma| n(J) = n\} } \frac{\lambda^n}{sym(J)} 
\int d\bfh\ \theta^*(\bfh_\Gamma) w(\bfh).
\ee
If the number of factors of $\psi$ in (\ref{order_n}) is odd then $n$ atoms and $\Gamma$ together
have an odd number of nodes and so cannot be matched up to form a 2-complex bounded by $\Gamma$.
(\ref{J_sum}) therefore covers this case as well, because the integral (\ref{order_n}) is zero while
(\ref{J_sum}) is zero because $\{J \in \gothA_\Gamma| n(J) = n\}$ is empty. The theorem is thus 
established. \QED

Note that we could have restricted $\psi(h_1,...,h_4)$ to be invariant under the 
gauge transformation
\be	\label{left_inv}
h_n \mapsto g h_n
\ee
without affecting the result. If one writes $\psi = \bar{\psi} + \Dg \psi$ where 
$\bar{\psi} = \int_G dg\ \psi(g\bfh)$, the gauge average of $\psi$, then one finds 
$I_0[\psi] = I_0[\bar{\psi}] + I_0[\Dg\psi]$ while $\calV[\psi] = \calV[\bar{\psi}]$
and $\Theta[\psi] = \Theta[\bar{\psi}]$ because $V$ and $\theta$ are gauge invariant.
It follows that the integral over $\Dg\psi$ cancels between numerator and denominator 
in the expectation value
\be		\label{expectation}
\langle \Theta^* \rangle = \int\calD\psi\ e^{-I_0[\psi] + \lambda \calV[\psi]} \Theta^*[\psi]/
\int\calD\psi\ e^{-I_0[\psi]}
\ee
and therefore this expectation value is unchanged if $\psi$ is replaced by $\bar{\psi}$, i.e.
is restricted to functions invariant under (\ref{left_inv}). (In (\ref{VEV}) the denominator of
(\ref{expectation}) was set to 1. We may simultaneously set $\int\calD\bar{\psi}
\ e^{-I_0[\bar{\psi}]} = 1$ and leave out the denominator also in the $\bar{\psi}$ formulation).
In the perturbation series of the $\bar{\psi}$ field theory the propagator (\ref{propagator}) 
is replaced by one that requires the connections to match only up to gauge. However, the gauge 
invariance of
$V$ and $\theta$ imply that the integrals over the connection can be evaluated in a gauge in which 
the connections on the atomic boundaries and $\Gamma$ really do match. The integrals over the 
new gauge degrees of freedom at nodes then simply contribute factors of 1 because the normalized 
Haar measure is being used.

\subsection{An example: The Turaev-Ooguri-Crane-Yetter model of BF theory}

A very simple gauge theory to which we can apply our scheme is BF theory. In this theory
the action for the boundary value problem in which the connection is fixed on the boundary
is $\int tr[ B\wedge F]$ where $F$ is the curvature of the connection, $B$ is a 2-form
which takes values in the Lie algebra of the gauge group $G$, and the trace is taken in this 
Lie algebra. Extremization of this action with respect to $B$ requires $F = 0$, that is, a flat
connection and in particular a flat connection on the boundary. Indeed a naive path integral 
quantization of this problem gives as the amplitude for the boundary connection a delta 
distribution with support on flat connections. It is therefore natural to discretize BF theory
as a simplicial local lattice gauge theory in which the amplitude for the connection on the 
boundary of each simplex is a delta distribution with support on flat connections or, more
concretely, $V$ is a delta distribution with support on flat connections on $\Gamma_5$.
Here flatness on $\Gamma_5$ means that the holonomy around any closed loop of
$\Gamma_5$ is trivial. The only such connections are the trivial connection and its gauge 
transforms, for which
\be
h_i{}^j = p_i q^{ij}
\ee
with $p_i \in G$ characterizing the gauge at the 4-valent node $i$ while $q^{ij} = q^{ji} \in G$
does so at the bivalent node in the chain connecting $i$ and $j$. Thus
\bearr
V_{BF}(h_i{}^j [h_j{}^i]^{-1}) & = & 
\prod_k \int dp_k \prod_{i<j} \dg(h_i{}^j [h_j{}^i]^{-1},p_i p^{-1}_j)		\\
& = & \prod_k \int dp_k \prod_{l<m} \int dq^{lm}\ \prod_{i\neq j} \dg(h_i{}^j, p_i q^{ij}).
\eearr
Substituting $V_{BF}$ into the formula (\ref{calV_def}) for the interaction term in the field 
action we find
\bearr
\calV_{BF}[\psi] & = & \frac{1}{5!} \int_{G^{20}} d^{20}h\ V_{BF}(h_i{}^j [h_j{}^i]^{-1})
\:\psi(h_1{}^i)\cdot ... \cdot \psi(h_5{}^i)			\\
& = & \frac{1}{5!}\prod_k \int dp_k \prod_{l<m} \int dq^{lm}\ \psi(p_1 q^{1j})\cdot ... \cdot
\psi(p_5 q^{5j})			\\
& = & \frac{1}{5!} \prod_{l<m} \int dq^{lm}\ \bar{\psi}(q^{1j})\cdot ... \cdot
\bar{\psi}(q^{5j}).
\eearr
$I_0[\bar{\psi}] - \lambda \calV_{BF}[\bar{\psi}]$ is precisely Ooguri's 
\cite{Ooguri} action for generating BF theory summed over 2-complexes.

\section{Spin foam formulation}	\label{spin_foam}

\subsection{From local lattice gauge theories to spin foams}

The results of the previous section can also be understood within the spin 
foam framework. This has the advantage, among others, of replacing the 
functional integrals of (\ref{VEV}) by more concrete multiple integrals 
over infinite sequences of real numbers, suggesting avenues 
toward regulating these integrals.

Any local lattice gauge theory with compact gauge group $G$ can be given a 
spin foam formulation. The spin foam formulation is conjugate to the 
connection formulation in the sense that if a connection is a history of the 
configuration variables of the model a spin foam is essentially a history of 
the momentum variables - the states in the spin foam picture are a sort of 
Fourier transform of the states in the connection picture. 

The spin foam sum for the amplitude
\be		\label{overlap}
\int d\bfh\ \theta^*(\bfh_\Gamma) w(\bfh),
\ee
of a boundary state $\theta$ on a fixed admissible 2-complex is obtained 
by expanding $V(\bfh_x)$ for each atom and $\theta(\bfh_\Gamma)$ on a basis of
``spin network functions" \cite{spinnet,Rei94}, and then integrating out the 
connection in each term of the resulting expansion of $\theta^* w$. Each 
non-zero term in this sum for (\ref{overlap}) is associated to a spin foam.
The translation of integrals over spacetime connections to sums over spin 
foams is described in detail in \cite{Rei97b}. Here we only outline its main 
features. They key necessary element for any reformulation of this type is an 
orthonormal basis of the space of distributions on the gauge group 
manifold. Since $G$ is assumed to be compact the Peter-Weyl theorem assures
us that the matrix elements of the unitary irreducible 
representations,\footnote{
Only one irreducible representation is selected from each class of unitarily
equivalent representations. The basis of functions on $G$ consists of the 
matrix elements of these selected representations.}
$U^{(r)}{}_m{}^n(g)$ (with $g$ ranging over $G$, $r$ labeling the 
representation, and $(m,n)$ the matrix elements) form an orthogonal basis of 
distributions. Indeed, since
\be		\label{orthogonality_1}
\int_G dg\ [U^{(r_1)}{}_{m_1}{}^{n_1}(g)]^*\: U^{(r_2)}{}_{m_2}{}^{n_2}(g)
= \frac{1}{dim\,r_1} \dg_{r_1 r_2} \dg^{m_1}_{m_2} \dg^{n_2}_{n_1}
\ee
with $dim\,r$ the dimensionality of the representation $r$, the set
$\{\sqrt{dim\,r}\: U^{(r)}{}_m{}^n\}_{r,m,n}$ is an orthonormal basis.

%

Any function of the parallel propagators along the edges of a graph may thus 
be expanded in terms of tensor products of representation matrices of the 
parallel propagators. Notice that a representation matrix of the parallel 
transporter along an edge is a two point tensor, with one index living
at each end of the edge. The subspace of functions that are gauge invariant 
at a given vertex is thus spanned by functions obtained from the tensor 
product of representation matrices by contracting the indices at that vertex 
with an invariant tensor, also called an ``intertwiner". (An example of an 
intertwiner when $G = SU(2)$ is the antisymmetric tensor $\eg_{ijk}$, which
corresponds to a trivalent vertex with each incident edge carrying the spin
1 representation).
An orthonormal basis of fully gauge invariant functions can be found by first 
selecting at each vertex an orthonormal basis $\{W^\bfr_I\}_I$ of the space 
of intertwiners for each set $\bfr$
of incident representations that admits intertwiners\footnote{
That is to say, each $\bfr = [r_1, r_2, ..., r_v]$ such that 
$r_1\otimes ... \otimes r_v$ contains the trivial representation, so that 
non-zero invariant tensors exist.}
and then forming the basis functions by contracting factors 
$\sqrt{dim\,r_e}\ U^{(r_e)}{}_m{}^n(g_e)$ for the edges $e$ with intertwiners 
belonging to the chosen intertwiner bases at the nodes. The data labeling a 
basis function, namely the representation on each edge and the basis 
intertwiner at each node, define a ``spin network" on the graph,\footnote{
Strictly the spin network consists of the subgraph carrying carrying 
non-trivial representations - edges that carry trivial representations are 
left off - together with the ``colouring" data consisting of the non-trivial 
representations and the intertwiners. In our present context of
lattice gauge theory this point is of no importance but in the continuum in 
which there is no convenient graph of all possible edges it is essential.}
and the basis functions are called ``spin network functions".

Now consider expanding $V(\bfg)$ into a sum of spin network basis functions.
\be		\label{V_expansion}
V(\bfg) = \sum_{\bfr_{..}, \bfI_{.}} A^{(\bfr_{..})\,\bfI_{.}} [\prod_{i<j} 
\sqrt{dim\,r_{ij}}]\phi^{(\bfr_{..})}_{ \bfI_{.}}(\bfg).
\ee
(Recall $g_{ij} = h_i{}^j [h_j{}^i]^{-1}$). 
$\phi^{(\bfr_{..})}_{ \bfI_{.}}(\bfg)$ is the (normalized) basis spin network 
function on $\Gamma_5$ with intertwiners $W_{I_i}$ at the 4-valent nodes and 
representations $r_{ij}$ on the edges connecting the 4-valent nodes.
%
%
\be
\phi^{(\bfr_{..})}_{ \bfI_{.}}(\bfg) = \prod_{i<j} \sqrt{dim\,r_{ij}}\: 
U^{(r_{ij})}{}_{m_{ij}}{}^{n_{ij}}(g_{ij})
\prod_k W_{I_k\,n_{1k}...n_{k-1\,k}}{}^{m_{k\,k+1}...m_{k5}}.
\ee
$\bf r_{..}$ denotes the whole matrix of representations $r_{ij}$ and 
$\bfI_{.}$ similarly denotes the sequence of basis intertwiners $I_i$.
The factor $[\prod_{i<j} \sqrt{dim\,r_{ij}}]$ in (\ref{V_expansion}) has been 
separated from the coefficient $A^{(\bfr_{..})\,\bfI_{.}}$ for later 
convenience.

Each spin network on the boundary of an atom defines a spin foam on the body 
of the atom: The intertwiner $I_i$ on the 4-valent node $i$ is the spin foam 
intertwiner on the spoke connecting the node with the 0-cell 
at the center of the atom; The representation $r_{ij}$ on the chain 
connecting nodes $i$ and $j$ defines the representation on the wedge
of the atom bounded by the chain.

The orthogonality relation (\ref{orthogonality_1}) and the orthogonality of 
the basis intertwiners then ensure that when $\theta^* w$ is integrated over 
the connection the only terms that remain are those in which these atomic spin
foams match up, in the sense that the intertwiners on the two halves of a 
1-cell of $J$ agree and the representations on all the wedges of a 2-cell 
agree.\footnote{
They must agree in the sense that if two wedges of a 2-cell have matching 
(``coherent") orientations then they must carry the same representation, 
whereas if they have opposite orientations they must have complex conjugate 
representations. These representations will be discussed more a little further
on.}
Moreover, they ensure that the intertwiners and representations match at the 
boundary $\Gamma$ where the spin foam meets the spin network in the expansion 
of $\theta$.

The weight, or amplitude, of each spin foam $S$ in the sum for (\ref{overlap})
that remains once the connection is integrated out is\footnote{
This formula should be viewed as somewhat schematic some details have
been ignored. The matching of basis intertwiners on coincident nodes is not 
between basis intertwiners in the same invariant tensor space but rather in 
dual spaces. If we use a raised index $I$ to label the dual basis 
intertwiners we see that the matching of intertwiner indices $A$ in the 
product (\ref{spin_foam_weight}) should always be between a downstairs and an 
upstairs index. For many groups one may choose the intertwiner bases so that
raising or lowering the intertwiner index is trivial, leaving $A$ unchanged,
but we are not sure that this is always true. 
Sign factors, not included in our formula, are also present for ``psuedo-real''
representations that some groups have which depend on the relative 
orientations of the coincident boundary edges on which neighboring atoms meet.
The proper resolution of these issues is explained in outline at the end of 
this section.}
\be	\label{spin_foam_weight}
c_{\di S}^* \prod_{x\in\,\mbox{\scriptsize 0-cells of}\,J} 
A^{(\bfr_{..})\,\bfI_{.}} 
\prod_{y \in\,\mbox{\scriptsize 2-cells of}\,J} dim\,r_y
\ee
where $c_{\di S}$ proportional to the coefficient of the spin network 
$\di S$ (the boundary of the spin foam) in the expansion of $\theta$, being
obtained from the coefficient by dividing by a factor of $\sqrt{dim\,r}$ for 
each edge of $\Gamma$.

\subsection{Divergences, regularization, and renormalization}

Spin foam models can have divergences when $J$ has closed 2-surfaces. 
For instance, in the model of BF theory that we have discussed the weight 
$w$ of a connection has redundant delta distributions, and is thus 
infinite, whenever $J$ contains topological 2-spheres. Other closed 
surfaces can also contribute singular factors\footnote{
The singular factor associated with a closed surface in BF theory
is
\be
\sum_{r\in R} (dim\,r)^\chi + \sum_{r\in PR} (- dim\,r)^\chi
+ \kappa\sum_{r\in C} (dim\,r)^\chi
\ee
where $\chi$ is the Euler characteristic
of the surface, $\kappa = 1$ if the surface is orientable and $0$ if it
is not, $R$ is the set of (unitary equivalence classes of) 
real irreducible representations, $PR$ is the set of psuedo-real 
irreps, and $C$ is the set of complex irreps (for definitions see 
\cite{Georgi}). When $G = U(1)$ we obtain a singular factor for every
orientable closed surface, whereas for $G = SU(2)$ only the four surfaces with
Euler characteristic $\chi \geq 0$ create divergences.} 

In the spin foam sum the infinities take the form of
a divergence of the sum over representations and can thus be regulated by 
somehow cutting off this sum. This can be accomplished by replacing $G$ with 
the quantum group $G_q$ with $q \neq 1$ a root of unity, which has a finite 
set of inequivalent irreducible representations. The spin foams are coloured 
with the representations and intertwiners of $G_q$, and $A$ is replaced by its 
natural generalization to such representations \cite{Turaev}\cite{CY} yielding
a finite sum. One can also simply cut off the sum, leaving the model otherwise 
unchanged. 

In lattice models of topological theories, or Yang-Mills theories for that
matter, the continuum limit theory is usually defined not by a sum  
over lattices but by the limiting values of the observables of the theory
as the lattice is refined. In the case of simplicial models the existence
of this limit requires that the model be independent of the triangulation used
(at least so long as the triangulation is very fine, i.e. has very many 
simplices). This requires that the regulated model be renormalized, because 
the number of divergences, and thus the number of the large, regulator 
dependent factors in the transition amplitudes, depends on the simplicial 
complex used. For BF theory on a triangulated manifold this 
means dividing out a factor $\dg(0)_{\mbox{\scriptsize regulated}} = 
[\sum_r (dim\,r)^2]_{\mbox{\scriptsize regulated}}$ for each independent 
2-sphere in $J$.\footnote{
The expansion on the basis of matrix elements of the Haar measure delta 
distribution on $G$ is 
\be
\dg(g) = \sum_r dim\,r\ tr\,U^{(r)}.
\ee}
This is the origin of the factor 
$\dg(0)_{\mbox{\scriptsize regulated}}^{N_0 - N_1}$ (with $N_0$ the number
of 0-{\em simplices} and $N_1$ the number of 1-{\em simplices} in the
triangulation) in the weight of a history in simplicial BF theory \cite{CY}.
Renormalization is more subtle in the context of a general admissible 2-complex
because the spheres do not necessarily span all the closed 2-surfaces (for
instance one can make a 2-complex having a torus as it's only closed 
surface). As far as we know the renormalization of the model has not been 
carried out in this wider context. 

In the present paper we are of course advocating a different approach to 
the continuum limit, in which one sums over 2-complex spacetimes. We have
not studied the issue of renormalization in this approach.

\subsection{The field theory in terms of spin foams}

Now let's see how the field theory (\ref{action_def}) can be expressed in 
terms of spin foams and networks. While this is qualitatively quite 
straightforward it turns out to be a little tricky in detail. The difficulty
has to do with the fact that the spin network basis is not quite unique.
Some conventions have to be introduced to define a particular basis.
When translating from an integral over connections to a sum over spin foams
on a particular, given, 2-complex one can choose the spin network 
bases on the atomic boundaries so as to simplify the calculations on that 
particular 2-complex.
On the other hand, in order to express our field theory in spin foam terms,
which boils down to expressing it in terms of the coefficients in spin network
expansions of $V$, $\theta$, and $\psi$, we have to choose a particular
spin network basis on the prototypical atomic boundary $\Gamma_5$. This
is then the basis on {\em every} atomic boundary in the complexes generated
by the field theory. The resulting inflexibility in the choice of basis
complicates the calculation of the amplitudes of Feynman diagrams.

We have taken a compromise route. We present the field theory
in spin foam language using a basis that makes the presentation as simple
and symetrical as possible, obtaining a very neat result. Then we outline 
how the amplitudes of the Feynman
diagrams can be calculated by exploiting the fact that one is free to change
to adapted spin network bases independently for each Feynman diagram, making 
the calculation as simple as in a spin foam model on a fixed
2-complex spacetime. 

It may well be that the results could be presented more cleanly using
a slightly more abstract ``basis independent'' approach to the spin network 
basis, in which the conventional choices that fix a particular basis are not 
made. Here we will stick with definite bases.

Let us begin the translation of the field theory by expanding $\psi$ using the 
Peter-Weyl theorem:
\be			\label{psi_expansion}
\psi(\bfh) = \sum_{\bfr} b^{(\bfr)\bfm}{}_{\bfn}\prod_{t=1}^4 \sqrt{dim\,r_t}
\ U^{(r_t)}{}_{m_t}{}^{n_t}(h_t).
\ee
The symmetry of $\psi$ under permutations of its arguments requires the 
coefficients $b$ to be similarly symmetric:
\be
b^{(\bfr)\,\bfm}{}_{\bfn} = b^{(\sg[\bfr])\,\sg[\bfm]}{}_{\sg[\bfn]}\ \ 
\forall \sg\in S_4.
\ee
The free action is easily expressed in terms of these coefficients if we write 
it as $I_0[\psi] = \frac{1}{2\cdot 4!} \int_{G^4} \psi^*\psi\ d^4h$. This is 
valid because $\psi$ is real. We find, using (\ref{orthogonality_1}), that
\be		\label{I_0_psi_expansion}
I_0[\psi] = \frac{1}{2\cdot 4!} \sum_{\bfr} [b^{(\bfr)\bfm}{}_{\bfn}]^* 
b^{(\bfr)\bfm}{}_{\bfn}.
\ee

The corresponding propagator is\footnote{
It can be evaluated by translating the field propagator 
$\langle \psi(\bfh)\psi(\bfh')\rangle_0$ or directly from 
(\ref{I_0_psi_expansion}). In carrying out the Gaussian integral 
in the second approach the reality conditions satisfied by $b$ (spelled out 
further on in the text) must be taken into account.}
\be		\label{b_prop}
\langle b^{(\bfr_1)\bfm_1}{}_{\bfn_1} [b^{(\bfr_2)\bfm_2}{}_{\bfn_2}]^*
\rangle_0
 =  \sum_{\sg\in S_4} \dg_{\bfr_1\sg[\bfr_2]} \dg^{\bfm_1}_{\sg[\bfm_2]}
\dg_{\bfn_1}^{\sg[\bfn_2]}		
\ee
with $\dg_\bfa^\bfb = \prod_{t=1}^4 \dg_{a_t}^{b_t}$.


To express the interaction term $\calV$ in terms of the coefficients $b$ in as 
clean a way as possible we shall introduce a further representation theoretic 
tool. If $g_{12} \in G$ defines parallel transport on the edge from node 1 to 
node 2 then $g_{21} = [g_{12}]^{-1}$ defines parallel transport along the 
inverse edge from 2 to 1. A unitary representation matrix 
$U^{(r)}_{m_1}{}^{m_2}(g_{12})$ can be expressed in terms
of $g_{21}$ as $[U^{(r)}(g_{21})]^{-1}_{m_1}{}^{m_2} = 
[U^{(r)}_{m_2}{}^{m_1}(g_{21})]^*$. It is just the parallel propagator from 
2 to 1 in the complex conjugate of the representation $r$.
Thus a spin network function of the parallel transporters along the oriented 
edges of a graph can be expressed just as well in terms of parallel 
transporters of the edges with different orientations, provided the 
representations on the reversed edges are replaced by their complex conjugates.

Now recall that in a spin network basis the representations $U^{(r)}$ that may
be placed on edges are particular representatives chosen one out of each 
unitary equivalence class. Moreover, in order to keep the definition of the 
spin network basis simple we shall use the same set of representatives on all 
edges. In general the conjugate representation $[U^{(r)}]^*$ will not be the 
representative $U^{(r^*)}$ of its own equivalence class, it will only be 
unitarily equivalent to it. For instance, all representations of $SU(2)$
are equivalent to their conjugates, but unitary irreducible representations of
non-integer spin are not real and so are not {\em equal} to their conjugates. 

We therefore introduce $\vareg^{(r)}_{mn}$, a unitary matrix such that
\be		\label{metric_def}
[U^{(r)}]^* = \vareg^{(r)\,\dag} U^{(r^*)} \vareg^{(r)}
\ee
and let $\vareg^{(r)\,mn} = [\vareg^{(r)}_{mn}]^*$, which is the inverse of 
$\vareg^{(r)}_{mn}$ in the sense that 
$\vareg^{(r)}_{mx}\vareg^{(r)\,nx} = \dg_m^n$. A few notes on $\vareg^{(r)}$:

\noindent $\bullet$ It follows from (\ref{metric_def}) that $\vareg^{(r)}_{mn}$
is an invariant tensor (an intertwiner) with $m$ a covector index of the 
representation $r^*$ and $n$ a covector index of the representation $r$. 

\noindent $\bullet$ (\ref{metric_def}) determines $\vareg^{(r)}$ up to phase. 
Consequently $\vareg^{(r^*)\,mn}$ equals $\vareg^{(r)\,nm}$ up to a phase 
factor, and thus when $r = r^*$, $\vareg^{(r)}$ is either purely 
symmetric or antisymmetric.\footnote{
For example consider the spin $1/2$ and spin $1$ representations of $SU(2)$ in
the standard eigenbasis of the generator $J_z$. In the spin $1/2$ 
representation $\vareg^{(\frac{1}{2})}_{mn} = \eg_{mn}$, while in the spin 1 
representation $\vareg^{(1)}_{mn} = \dg_{m,-n}$, so it is antisymmetric for 
spin $1/2$ and symmetric for spin 1. Indeed one finds that $\vareg^{(j)}_{mn}$
is symmetric for all integer $j$ and antisymmetric for all half odd integer 
$j$.}
When $r \neq r^*$ one is of course free to choose the representative of the 
equivalence class of the conjugate representation as one likes. One 
could simply choose the conjugate representation itself and use 
$\vareg^{(r)}_{mn} = \dg_{mn}$. We are therefore free to set 
$\vareg^{(r)} = \vareg^{(r^*)}$ for all $r$.\footnote{
When $r = r^*$ and $\vareg^{(r)}$ is symmetric the representation $r$ is real;
If $r = r^*$ and $\vareg^{(r)}$ is antisymmetric $r$ is psuedo-real; Finally, 
if $r \neq r*$ $r$ is complex. See \cite{Georgi}.}
\noindent $\bullet$ $\vareg^{(r)}$ can be used as a (possibly antisymmetric) 
metric to turn a representation $r$ vector (carrying an upstairs index) into a
representation $r^*$ covector (carrying a downstairs index) according 
$a^{(r^*)}_m = \vareg^{(r)}_{mx} a^{(r)\,x}$. The inverse, index raising, 
operation is $a^{(r)\,x} = \vareg^{(r)\,xm} a^{(r^*)}_x$. Of course complex 
conjugation also turns unitarily transforming $r$ vectors into $r^*$ covectors
and {\em vice versa}. The index positioning in all our equations is consistent
with this fact.

The relation (\ref{metric_def}) can be used to define a modified, more 
symmetric spin network expansion of $V$ which depends only minimally on the 
orientations chosen for the edges of $\Gamma_5$. Notice that
\bearr
U^{(r)}_{m_1}{}^n (h_1 h_2^{-1}) & = & U^{(r)}_{m_1}{}^x(h_1)\: [U^{(r)}_n{}^x 
(h_2)]^*\\
	& = & U^{(r)}_{m_1}{}^x (h_1)\: \vareg^{(r)}_{xy}\: U^{(r^*)}_{m_2}{}^y
(h_2)\: \vareg^{(r)\,n m_2}.
\eearr
Thus adopting the notation $r_{ij} = r^*_{ji}$ for $i>j$ we may write
\be		\label{V_expansion2}
V = \sum_{\bfr_{..}, \bfI_{.}} A^{(\bfr_{..})\,\bfI_{.}} 
[\prod_{i\neq j} \sqrt{dim\,r_{ij}}]
U^{(r_{ij})}{}_{m_{ij}}{}^{x_{ij}}(h_i{}^j)
\prod_k W_{I_k}^{m_{k1}...m_{k\,k-1}m_{k\,k+1}...m_{k5}}\prod_{i<j} 
\vareg^{(r_{ij})}_{x_{ij}x_{ji}}.
\ee
The lower indices of the intertwiners appearing in (\ref{V_expansion}) have 
been raised using $\vareg$. The invariance of $\vareg$ implies that the 
resulting tensors $W_{I_k}^{\bfm_{k.}}$ are also 
invariant and thus intertwiners for the incident representations $r_{ki}$. It 
is easy to verify that they form an orthonormal basis of such intertwiners. 
(\ref{V_expansion2}) is thus a spin network expansion of $V$ with the
bivalent nodes assigned the intertwiners $\vareg^{(r_{ij})}_{x_{ij}x_{ji}}$ 
(which is $\sqrt{dim\,r_{ij}}$ times the unique (mod phase) normalized bivalent
intertwiner). In (\ref{V_expansion2}) the orientations of the edges manifest 
themselves only in the ordering of the indices in these $\vareg$s. 

Evaluating $\calV$ is now straightforward. If we use the reality of $\psi$ to 
replace $\psi$ by $\psi^*$ in (\ref{calV_def}) the integrals can be carried 
out using (\ref{orthogonality_1}) and we obtain
\be 	\label{spinfoam_calV}
\calV = \frac{1}{5!}\sum_{\bfr_{..}, \bfI_{.}} A^{(\bfr_{..})\,\bfI_{.}} 
\prod_k \:
[b^{(\bfr_{k\cdot})\,\bfm_{k\cdot}}{}_{\bfx_{k\cdot}}]^*\:
W_{I_k}^{\bfm_{k\cdot}}
\ \prod_{i<j} \vareg^{(r_{ij})}_{x_{ij}x_{ji}}.
\ee 

$\calV$ is even simpler when expressed in terms of the spin network 
coefficients in an expansion of the ``gauge invariant" field 
$\bar{\psi}(\bfh) = \int_G \psi(g\bfh)\ dg$. A spin network
type expansion of this field can be obtained by inserting into the expansion 
(\ref{psi_expansion}) 
of $\psi$ the projector
\be
\int_G \prod_{t=1}^4 \ U^{(r_t)}{}_{m_t}{}^{n_t}(g) dg = \sum_I 
[W_I^{(\bfr) m_1 m_2 m_3 m_4}]^* W_I^{(\bfr) n_1 n_2 n_3 n_4}
\ee
onto invariant $r_1\otimes r_2 \otimes r_3 \otimes r_4$ tensors. Thus 
\be			\label{psibar_expansion}
\bar{\psi}(\bfh) = \sum_{\bfr,I} c^{(\bfr)I}{}_{\bfn} W_I^{(\bfr)\,\bfm} 
\prod_{t=1}^4 \sqrt{dim\,r_t}\ U^{(r_t)}{}_{m_t}{}^{n_t}(h_t)
\ee
with $c^{(\bfr)I}{}_{\bfn} = b^{(\bfr)\bfm}{}_{\bfn} [W_I^{(\bfr)\,\bfm}]^*$.
In terms of the coefficients $c$ the interaction term is
\be 	\label{spinfoam_calV2}
\calV = \frac{1}{5!}\sum_{\bfr_{..}, \bfI_{.}} A^{(\bfr_{..})\,\bfI_{.}} 
\prod_k\: [c^{(\bfr_{k\cdot})\, I_k}{}_{\bfx_{k\cdot}}]^*\  
\prod_{i<j} \vareg^{(r_{ij})}_{x_{ij}x_{ji}},
\ee 
It's just $A$ times the result of contracting together the indices of five 
$c^*$ in the pattern of a 4-simplex. More precisely it can be obtained from 
(\ref{V_expansion2}) by replacing the intertwiners $W_I$ by $[c^I]^*$ and 
setting all the parallel transporters to $\One$. This prescription can
also be applied to $\theta$ to express $\Theta$ in terms of the coefficiants 
$c^*$.

$\vareg$ also allows us to express the reality conditions satisfied by the 
coefficients $b$:
$\psi^* = \psi$ and (\ref{metric_def}) implies that
\be		\label{reality}
b^{(\bfr^*)\bfm}{}_{\bfn} = [b^{(\bfr)\bfx}{}_{\bfy}]^* 
\vareg^{(\bfr)\,\bfm\bfx}
\vareg^{(\bfr)}_{\:\bfn\bfy}
\ee
(with $\vareg^{(\bfr)\,\bfm\bfn} = \prod_{t=1}^4 \vareg^{(r_t)\,m_t n_t}$).

We are now in a position to rewrite the functional integral (\ref{VEV}) as an 
ordinary integral over the coefficients $b$ (which are finite in number if 
the spin foam model has been regulated by cutting off the sum over 
representations). Moreover we may evaluate the Feynman diagrams of the 
perturbation series in $\lambda$ using the propagator
(\ref{b_prop}) and the reality conditions (\ref{reality}).

What we have given is just about the simplest, most symmetric statement of 
the spin foam formulation of the theory. However, it is not the most 
convenient form for actually evaluating the Feynman diagrams. In fact it is 
best to first select the Feynman diagram one wishes to evaluate, 
corresponding to a particular 2-complex $J$, and {\em then} choose the spin 
network basis for expanding $\Theta^*$ and the factor of $\calV$ 
corresponding to each atom. To define this adapted set of spin network bases 
we choose an orientation for each 1-cell and 2-cell of $J$. The orientations 
on the 2-cells induce an orientation on each edge of $\Gamma$ and each atomic 
boundary edge that matches the sense of circulation of the 2-cell that it cuts.
The orientations on the 1-cells define a sign on each node, which is positive 
if the 1-cell is outgoing at the node and negative if it is incoming. Now we 
choose spin network basis functions corresponding to these orientations of the
edges. We obtain an expansion of $V$, rather like that given in 
(\ref{V_expansion}) with coherent orientations on the chain of two edges 
connecting a pair of nodes. However, we do not necessarily have two outgoing 
and two incoming edges at each node as in (\ref{V_expansion}). The intertwiner 
bases are also adapted. At each positive node a basis $\{ W_I^{(\bfr)}\}$ with 
indices suitably lowered for incoming edges, is used. At the corresponding
negative node (the negative node that lies on the same 1-cell) the intertwiners
$[W_I^{(\sg[\bfr])}]^*$ are used. Here $\sg \in S_4$ is the mapping of the 
incident edges at the 
negative node to the corresponding incident edges at the positive node. These
complex conjugate and permuted intertwiners are still orthonormal and have 
the index positions compatible with our convention for the orientations of 
the edges.

With these conventions (and with the correspondingly adjusted coefficients $A$ 
and $c_{\di S}$) the amplitude of the Feynman diagram is relatively easy to 
obtain. The result is as described in (\ref{spin_foam_weight}): It is a sum 
over histories consisting of representations and basis intertwiners, 
with only histories in which every 2-cell $y$ carries a single representation 
$r_y$ - the common representation on the atomic boundary edges that cut the 
2-cell and on the edges of $\Gamma$ that bound it - and each 1-cell carries a 
single intertwiner basis index $I$, the common value of the indices at the 
positive and negative nodes on the 1-cell. Such a history is a spin foam. 
The amplitude for each history is the product of the coefficients
$A$ for the atoms times $c^*_{\di S}$ and a factor $dim\,r_y$ for each 2-cell. 
This last factor results from the factors of 
$\vareg^{(r_{ij})}_{x_{ij}x_{ji}}$ that appear in (\ref{spinfoam_calV}) 
and (\ref{spinfoam_calV2}), associated to each wedge. When the orientations 
of the edges are changed to match those of the 2-cells these are replaced
by Kronecker deltas. Then, when the propagators for $b$ are substituted in,
the Kronecker deltas associated to wedges in a given 2-cell end up being 
contracted in a chain around around the 2-cell, thus contributing a factor
$dim\,r_y$.  

\section{Some closing remarks}	\label{closing}

\subsection{On the possible finiteness of the field theory corresponding to
regulated spin foam models}

The number of admissible 2-complexes increases very rapidly with the number of
atoms. In fact the number of complexes of $n$ atoms with a given boundary 
$\Gamma$ having $m$ 4-valent nodes is given approximately by
\be
\sum_{\{J \in gothA_\Gamma| n(J) = n\}} = (\frac{4!}{2})^{(5n + m)/2} 
\frac{(5n + m)!}{([5n + m]/2)!}\frac{1}{n!(5!)^n}\frac{1}{sym(\Gamma)}.
\ee
This increases as $(n!)^{3/2}$ for large $n$, so unless the vertex function
$V$ is very special indeed the radius of convergence of the series 
(\ref{Feynman_sum}) is zero. 

What prospect is there of making sense of the functional integral (\ref{VEV})?
In fact it is not unreasonable to hope that once the spin foam model is 
regulated by cutting off the sums over representations (as we discussed in 
the context of BF theory) that the integral (\ref{VEV}) can be assigned a 
finite value by analytic continuation. We will suppose that the finite set 
of representations summed over in the cut off model includes the 
representation $r^*$ whenever it includes $r$. Then $\psi$ in the cut off 
model is a real function on $G^4$ determined by a finite set of parameters. 
These paramenters can be the $b^{(\bfr)\,\bfm}{}_\bfn$, but we will use a set 
of linear combinations $\{x_p\}_{p=1}^N$ of these chosen so that they are 
real and $I_0[\psi] = \bfx^T \bfx$. Then (\ref{VEV}) becomes
\be		\label{F}
\int_{\Real^N} d^N x\ e^{-\bfx^T \bfx}\: e^{\lambda\calV^{p_1...p_5} 
x_{p_1}\cdot ...\cdot x_{p_5}}
\ \Theta^{*\,q_1...q_m} x_{p_1}\cdot ...\cdot x_{p_m},
\ee
a function of $\lambda$ which we shall call $F(\lambda)$. For simplicity we 
will suppose that $V(\bfg)$ is real, as is the case in BF theory, and in the 
models of \cite{Rei97b} and \cite{BC}, then also $\calV^{n_1...n_5}$ is real 
and we see that the 
integral is convergent when $\lambda$ is pure imaginary - the 
integral is a Gaussian times a function of modulus $O(|x|^m)$. Can the result 
be analytically continued away from the imaginary axis?

To see that the answer is quite possibly ``yes'' consider the simplest analog 
of the integral (\ref{F}):
\be
f(\lambda) = \int_{-\infty}^\infty dx\: e^{-x^2} e^{\lambda x^5}.
\ee
The coefficients in the formal power series expansion of $f$ about 
$\lambda = 0$ also diverge as $(n!)^{3/2}$ for large $n$ and, like (\ref{F}) 
the integral converges for purely imaginary $\lambda$. Moreover, $f$ can be 
continued to the entire complex plane, with a fivefold branch cut extending 
from $0$ to $\infty$. To show this one first writes $f$ as
$f(\lambda) = h(\lambda) + h(-\lambda)$ where
\be
h(\lambda) = \int_0^\infty dx e^{-x^2} e^{\lambda x^5}.
\ee
This integral is convergent when $\Re \lambda \leq 0$. It may be reexpressed 
in terms of $\lambda^{-1/5}$ and the rescaled integration variable 
$y = \lambda^{1/5}x$. From this form one obtains a series for $h$ in powers 
of $\lambda^{-1/5}$ that is convergent for all finite values of this 
parameter, and the claimed result follows.

The first step in this argument can be repeated for the regulated functional 
integral (\ref{F}). Let $M$ be the union of rays in $\Real^4$ on which 
$\calV^{n_1...n_5}x_{n_1}\cdot...\cdot x_{n_5}
\leq 0$. Then
\be
F(\lambda) = H(\lambda) + H(-\lambda)
\ee
with 
\be
H(\lambda) = \int_M d^N x\ e^{-\bfx^T\bfx}\: e^{\lambda\calV(\bfx)} 
\ \Theta^*(\bfx).
\ee
Again we find that the integral converges for $\Re \lambda \leq 0$. The 
question is now whether it can be continued beyond this domain. It seems 
likely that this would be the case for generic models, especially 
as it is finite on $\Re \lambda = 0$. If it is continuable then $F(\lambda)$ 
will be defined and finite on an open region in the complex plane, possibly
including $\lambda = 1$ which is the value corresponding most closely
to a simple sum of the spin foam model over all admissible 2-complexes.

\subsection{Generalizations of the formalism}

Our formalism can easily be generalized to a wider class of 2-complexes.
We have allowed only atoms which are dual 2-skeletons of 4-simplices,
which means that they have five spokes each of which is four valent.
The field theory can be extended so that it generates 2-complexes including
atoms with any given number $p$ of 4-valent spokes (which are dual
to 4-polyhedra bounded by $p$ 3-simplices). All that is necessary is to
include a suitable interaction term in the action (\ref{action_def}) formed 
from
a vertex function for the new type of atom like $\calV$ was formed from
the vertex function $V$ and $\Theta$ from the boundary state $\theta$.
The expression (\ref{overlap}) for the amplitude of the state $\theta$ on the 
boundary $\Gamma$ of $J$ can be viewed as the partition function for
a closed 2-complex consisting of $J$ and one copy of a new type of atom
with boundary $-\Gamma$ ($\Gamma$ with reversed orientation) and vertex
function $\theta^*$. Adding $\Theta^*$ to the action would generate
all complexes consisting of both this new type of atom and the original type.
 
Even with the above generalization all atoms have spokes of valence 
four and thus all bounding graphs of atoms and of complexes have non-trivial
nodes of valence four only. The theory may be generalized to acomodate other
valences $q$ by introducing additional fields depending on $q$ 
group elements instead of four like $\psi$ does, and adding a quadratic free 
action to the total action for each such field.

Any model having a finite set of types of atoms can be handled by these means,
including of course models with spacetimes of any dimensionality - to the 
extent that this dimensionality can be captured in the combinatorial 
structure of the atoms. 

\subsection{Matrix models}

When our formalism is applied to two dimensional models with atoms dual 
to triangles we find that the analog of $c^{(\bfr)\,I}{}_\bfn$ is
$c^{(r_1)}{}_{n_1 n_2} \equiv b^{(r_1,r_2)\,m_1 m_2}{}_{n_1 n_2}
\frac{1}{\sqrt{dim\,r_1}} \dg_{r_1 r_2^*} \vareg^{(r_1)}_{m_1 m_2}$
($r_1$ has to equal $r_2^*$ because $r_1\otimes r_2$ contains the trivial
representation only in this case. $\vareg^{(r)}/\sqrt{dim\,r}$ is the only
normalized bivalent intertwiner). $c^{(r)}$ is essentially the matrix of 
matrix models \cite{matrix}: The triangular matrix model is recovered by 
choosing the vertex function to be $V(g_{ij}) = A (dim\,r)^3/3!
tr\,U^{(r)}(g_{12}g_{23}g_{31})$ so that only one representation $r$
appears in its spin network expansion. Then $\calV = A/3! 
c^{(r)}_{n_1}{}^{n_2}c^{(r)}_{n_2}{}^{n_3}c^{(r)}_{n_3}{}^{n_1}$. 

\section*{Acknowledgements}

We thank Roberto DePietri for discussions and help, and M.R. would like to
thank Rodolfo Gambini for a discussion on the subject of this paper. This
work was partially supported by NSF grant PHY-95-15506, PPARC grant
PPA/6/s/1998/00613 and a gift from the Jesse Phillips Foundation.

\end{document}